\documentclass[%
 reprint,%
 amssymb, amsmath,%
 aip,rsi,%
]{revtex4-1}

\usepackage{docs}%
\usepackage{bm}%
\usepackage[colorlinks=true,linkcolor=blue]{hyperref}%
\expandafter\ifx\csname package@font\endcsname\relax\else
 \expandafter\expandafter
 \expandafter\usepackage
 \expandafter\expandafter
 \expandafter{\csname package@font\endcsname}%
\fi
\usepackage{graphicx}
\usepackage{enumitem}
\setdescription{leftmargin=*}

\begin{document}
\noindent

\title{Analysis of Granular Packing Structure by Scattering of THz Radiation}%

\author{Philip Born}%
\email{philip.born@dlr.de}
\affiliation{Institut f\"ur Materialphysik im Weltraum, Deutsches Zentrum f\"ur Luft- und Raumfahrt e.V., 51170 Cologne, Germany}%

\author{Karsten Holldack}
\affiliation{Institut f\"ur Methoden und Instrumentierung der Forschung mit Synchrotronstrahlung, Helmholtz-Zentrum Berlin f\"ur Materialien und Energie GmbH, Albert-Einstein-Str. 15, 12489 Berlin, Germany}%

\date{\today}%
\revised{tbd}%

\maketitle



\section*{abstract}
Scattering methods are widespread used to characterize the structure and constituents of matter on small length scales. This motivates this introductory text on identifying prospective approaches to scattering-based methods for granular media. A survey to light scattering by particles and particle ensembles is given. It is elaborated why the established scattering methods using X-rays and visible light cannot in general be transferred to granular media. Spectroscopic measurements using Terahertz radiation are highlighted as they to probe the scattering properties of granular media, which are sensitive to the packing structure. Experimental details to optimize spectrometer for measurements on granular media are discussed. We perform transmission measurements on static and agitated granular media using Fourier-transform spectroscopy at the THz beamline of the BessyII storage ring. The measurements demonstrate the potential to evaluate degrees of order in the media and to track transient structural states in agitated bulk granular media.


\section{Introduction}

Scattering is a highly efficient method to characterize the structure of matter at small distances. Prominent examples like X-ray diffraction (XRD), small-angle X-ray scattering (SAXS), small-angle light scattering (SALS) (also called laser light scattering) and static light scattering (SLS) rely on the angular redistribution of electromagnetic waves during their propagation within the matter of interest \cite{Glatter1995,Xu2002,Brown1996}. They all give space- and time-averaged information of atomic, molecular or colloidal form and structure factors. This is accompanied by moderate instrumental effort and short measurement times, as in the case of the visible light scattering techniques.

Scattering methods are so fundamental to materials science, that it is worth discussing their potential for studies on granular matter. Apparently, scattering methods could provide experimental benefits compared to imaging methods. A single exposure of an area detector or a detector array, e.g. a CCD camera, could reveal ensemble-averaged structural information on a sample. Thus, even investigation of non-stationary processes in bulk samples should become possible using scattering methods with good time resolution, which is a blind spot of investigations with imaging methods (compare the introductory article to the special topic section on imaging in this issue).

The direct transfer of established scattering approaches to granular media is not possible though. The mentioned scattering methods rely on resolving the momentum transfer to an incident wave by measuring the angular scattering pattern. We discuss in the following section fundamental situations of scattering of electromagnetic radiation from particles and particle ensembles. It becomes obvious that in most situations scattering from dense granular media is distinct from X-ray and light scattering in atomic or colloidal media as the mean free path of the radiation becomes very short. Light transport through granular samples yet happens by scattering, and the transport properties of a granular medium will reflect the scattering properties of the particle ensemble. We show that with Terahertz (THz) radiation this transport properties become sensitive to the packing structure. We intend a beginners guide to the relevant approaches and literature. A coherent derivation of scattering properties starting from individual particles ranging up to dense media is not possible. Fundamental regimes can be classified by the ratio of individual particle size to wavelength, which will persist in dense media. The discussion will thereby be restricted to elastic scattering, inelastic processes like Raman scattering and absorption processes will be neglected.

We then give an introduction to design and setup of experiments to probe scattering properties of granular media in section \ref{sec:THzscatt}. We finally show light transport measurements on granular media which undergo disorder-order transitions in section \ref{sec:experiments}. These demonstrate the potential to monitor structural changes in-situ.

Scattering-based methods for granular media thus do not provide the high levels of information as imaging methods do under optimal conditions. Size and shape distributions can be determined with imaging without a priori assumptions, and local structural properties can be probed without the need for spatial averaging. Still, scattering-based methods might find their place in granular matter science, as they allow for a time-resolution which for index-matching or tomography could be reached only with high instrumental effort. 


\section{Scattering basics}
\label{sec:basics}
\begin{figure*}
	\centering
		\includegraphics[width=0.8\textwidth]{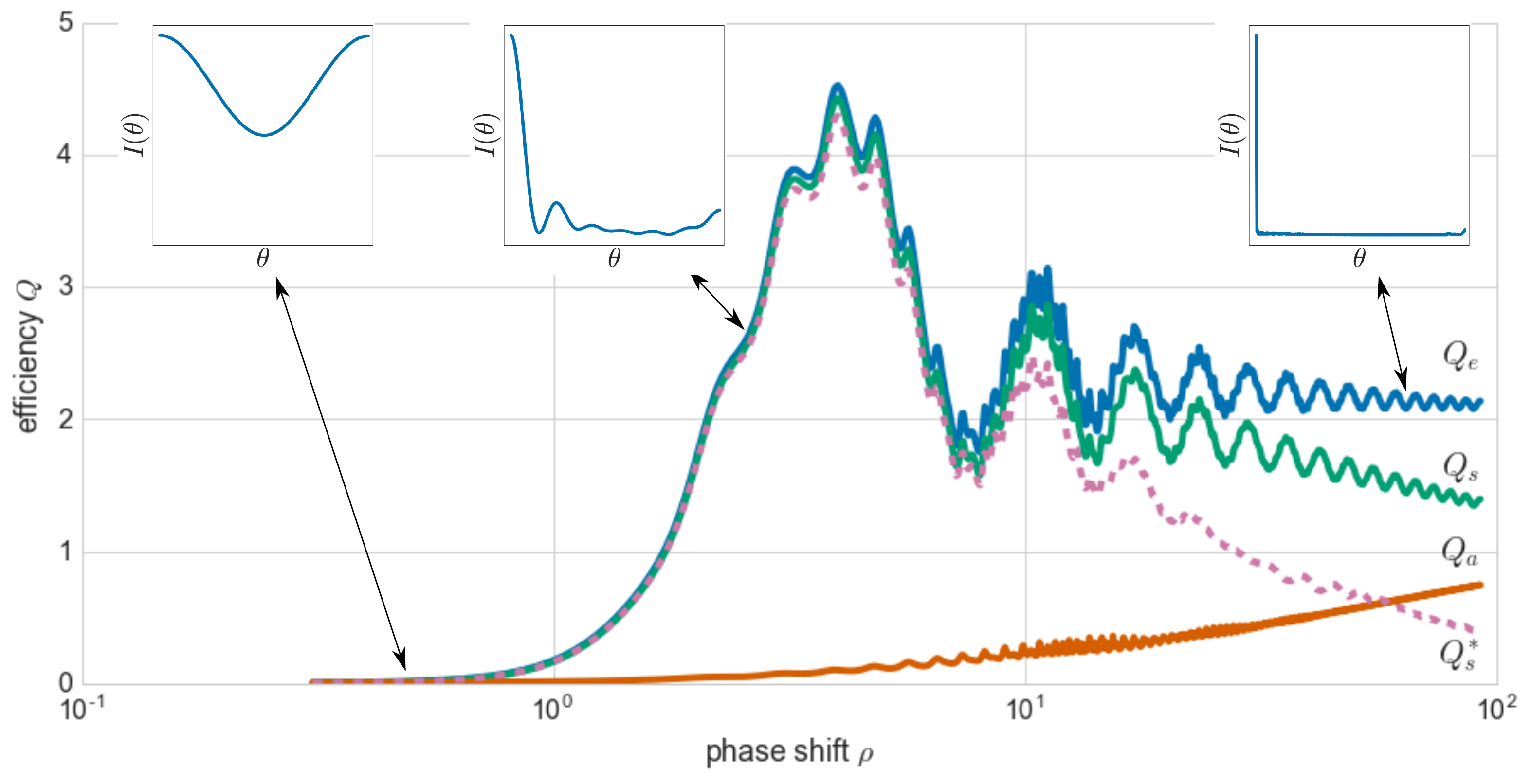}
	\caption{Theoretical expectation for the extinction ($Q_{e}$), scattering ($Q_{s}$) and absorption efficiency ($Q_{a}$) of a dielectric sphere. Three regimes are distinguishable from this plot: First, a regime with vanishing efficiencies and isotropic scattering (upper left inset) at $\rho < 1$. Second, a regime with $Q_{e}$ approaching 2 and $Q_{s}$ and $Q_{a}$ approaching 1 for $\rho\gg 1$. Scattering is confined to forward scattering (upper right inset). In between these regimes is a regime where the efficiencies and the scattering pattern depend strongly on $\rho$. The efficiencies and pattern are calculated for a sphere with a refractive index $m=1.59+i\cdot0.005$. The efficiencies are displayed as a function of $\rho$, the phase shift created by the particles (Eq.~\ref{eq:rho}). This allows to scale the efficiency values to different wavelengths, particle sizes and refractive indexes. $Q_{s}^{\ast}$ indicates the scattering efficiency that is measurable with a poorly collimated detector, which cannot discriminate between unscattered light and light scattered to small angles $\theta$ (dashed line).}
	\label{fig:Q}
\end{figure*}

We start the survey on scattering basics with single particle scattering properties, from which major regimes of scattering from particle ensembles can be deduced. Predictions on the light transport through granular media can be made from the scattering basics, which serves as a decision guide to select a scattering-based method for characterization. We finally derive in this section the relation among structure sensitivity and the wavelength of the used light. This short survey is designed only to highlight the approaches for scattering-based methods for granular media. Good references for scattering from particles \cite{Hulst1981,Bohren1983,Born1999} and particle ensembles \cite{Born1999,Ishimaru1999,Mishchenko2014} in general are available. 

\subsection{Scattering regimes for individual particles}
\label{sec:singleparts}

The elastic scattering of electromagnetic radiation by a particle can be classified using the extinction efficiency $Q_{e} = \sigma_{e}/(\pi a^{2}) = Q_{a}+Q_{s}$, the ratio of the extinction cross-section $\sigma_{e}$ to the geometrical cross-section of the particle with radius $a$, and the sum of the absorption and scattering efficiency $Q_{a}$ and $Q_{s}$. The $Q_{e}$ of spherical dielectric particles of different sizes can be scaled on top of each other using the average shift in phase 
\begin{equation}
\rho = 2 x |m-1|
\label{eq:rho}
\end{equation} 
of the transmitted wave acquired by passing the particles \cite{Hulst1981,Bohren1983,Mishchenko2014}. $x = 2\pi a /\lambda$ is the size parameter of a particle, $a$ is the particle radius, $\lambda$ the wavelength, and $m$ the refractive index of the particles. $m$ is determined relative to the surrounding medium and in general is complex, where the imaginary part describes the absorption properties of the particles. Analytical calculation of the efficiencies is generally only possible for spherical or cylindrical particles \cite{Bohren1983}, but numerical methods and high computing powers increasingly allow for calculations for other shapes \cite{Mishchenko2000}. Such a calculated master curve for $Q_{e}$ of spherical particles is given in Fig.~\ref{fig:Q}. Three regimes become specifiable:

\begin{description}
	\item[\emph{$\rho < 1,\ Q_{e} < 1$:}] The scattering by particles is rather isotropic (see left inset in Fig.~\ref{fig:Q}), and the particles scatter and absorb less radiation than is impinging on their geometrical cross-section in this regime. This regime is achieved either by particles smaller than the wavelength ($x$ small), or by close matching of the refractive indexes of the particle and the surrounding medium ($m$ small). This is the regime for imaging through the particles, as is achieved by using long-wave radiation to image objects in granular media (see the contribution by F. Ott, S. Herminghaus K. Huang in this issue), or by laser-sheet scanning through index-matched granular media (see the contribution by J. A. Dijksman and N. Brodu in this issue). Scattering in this regime can be described by the Rayleigh- or Rayleigh-Gans-Debye-approximations (RGD, see Sec.~\ref{sec:firstorder}).
	
	\item[\emph{$\rho \gg 1,\ Q_{e} \approx 2$}:] Scattering of light is confined into a narrow cone close to the forward direction (right inset in Fig.~\ref{fig:Q}). In this limit the particles are either much larger than the wavelength or having large refractive index mismatches. The particles theoretically extinct twice as much radiation than is impinging on their geometrical cross-section. However, this is only measurable with highly optimized setups like SALS-instruments where very small scattering angles can be discriminated. The light scattered close to the forward direction and the transmitted light cannot be discriminated with uncollimated detectors, and measured scattering efficiencies reduce (dashed line in Fig.~\ref{fig:Q}). This is the regime for conventional imaging of granular particles using visible light or for tomography using X-rays (see the contribution by S. Weis and M. Schr\"oter in this issue). Scattering in this regime can often be well approximated by the theory of Fraunhofer diffraction by opaque planar discs.
	
	\item[\emph{$\rho \geq 1,\ Q_{e}$ variable}:] The scattering patterns depend strongly on the particle size in this regime (central inset in Fig.~\ref{fig:Q}). Scattering resonances occur, such that efficiencies can exceed even 2. Scattering patterns and extinction efficiencies have to be calculated using the Lorenz-Mie theory for scattering \cite{Hulst1981,Bohren1983}, which is more involved than the approximations valid in the other regimes and requires knowledge of the complex refractive index. 
\end{description}

Particle size analysis by measuring angular scattering patterns is possible in all three scattering regimes. This is accomplished by fitting calculated scattering patterns to the measured patterns. This is particularly convenient in the approximation of Fraunhofer scattering, as no knowledge on the refractive index of the particles is required and the scattering efficiency is independent of particle size. In the regime of Rayleigh- and Rayleigh-Gans-Debye scattering, the strong dependence of the scattering efficiency on the particle size ($\propto a^{6}$ in the limit of Rayleigh scattering \cite{Hulst1981}) exacerbates deduction of size distributions from measured scattering patterns. In the intermediate regime, as no approximate formulas are available, knowledge of the complex refractive index of the particles is required for particle size analysis from measured scattering patterns. The nontrivial dependence of the scattering efficiency on the particle size in this intermediate regime additionally allows for particle size analysis by spectroscopic methods \cite{Born2015}. 

Analysis of packing structures, however, is not possible in all three regimes. A unique approach to structure analysis, like measuring the angular scattering pattern for particle size analysis, cannot be given. Two established approaches to analyze packing structures will be discussed in the following sections.

\subsection{Scattering from particle ensembles}
\label{sec:media}
\begin{figure*}
	\centering
		\includegraphics[width=0.8\textwidth]{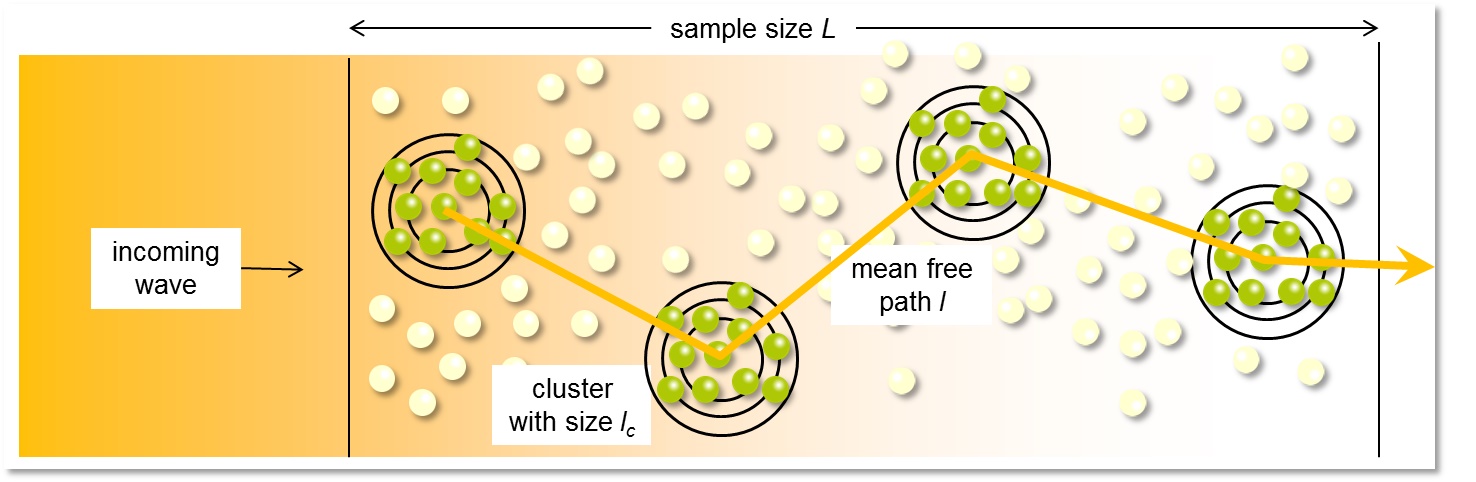}
	\caption{Scheme of light propagation in dense media. Light is incident onto a sample with thickness $L$ and gets scattered at certain particles. The particle positions around the scattering particle will be correlated up to a maximal correlation length $l_{c}$. This modulates the scattering pattern of the particle by interference. The light travels on average a distance $l$ before being scattered again. The ratios of $L$, $l$ and $l_{c}$ determine the transport properties of the media.}
	\label{fig:bulk}
\end{figure*}

Scattering from an ensemble of particles differs from scattering from individual particles due to the electromagnetic interaction among the particles and the interference of the scattered radiation \cite{Mishchenko2014}. The suitable approach for analyzing structures of the particle ensemble can be identified using the parameters $L$, $l$ and $l_{c}$, which are indicated in Fig.~\ref{fig:bulk}. $L$ denotes the geometric size of the sample. $l_{c}$ is a spatial correlation length in the sample. We define $l_{c}$ as any distance of a positive correlation peak in the pair correlation function $g(r)$ of the particle ensemble. The largest $l_{c}$ can become comparable to $L$ for a single-crystalline arrangement of the particles, and reduces to a few particle diameter for disordered hard-sphere packings \cite{Waseda1980,Hansen2005}. The mean free path $l$ of the electromagnetic radiation within the ensemble of particles gives an average length that the electromagnetic wave travels before being scattered again. $l$ is determined by the number density of the particles $\varphi$ and their scattering efficiency $Q_{s}$, 
\begin{equation}
	l = 1/(\varphi\cdot Q_{s}\cdot \pi a^{2}).
\label{eq:l}
\end{equation}
The scattering efficiency $Q_{s}$ of the particles in the sample is exactly the same as the one discussed in Sec.~\ref{sec:singleparts} only for dilute samples, and is generally reduced compared to the dilute value by various effects in dense ensembles \cite{Fraden1990,Auger2011}. Three major regimes of light scattering in particle ensembles become distinguishable by their $l$ and $L$ \cite{Ishimaru1999}: 

\begin{description}
	\item[\emph{$l>L$}:] The scattering length is larger than the sample size $L$ for optically dilute samples. This is the limit where the first-order approximation, as discussed in Sec.~\ref{sec:firstorder}, is valid for the whole particle ensemble. The structure factor, which describes the reciprocal structure of a particle ensemble can be obtained from the angular scattering pattern within the limits of the first-order approximation.
	
An optical dilution $l>L$ can be achieved in different ways (compare Eq.~\ref{eq:l}). Either the sample is very dilute and has a low number density $\varphi$. Or the particles have a very small scattering efficiency $Q_{s}$, which can be achieved by using wavelengths much larger than the particles or by index matching of the particles (see Sec.~\ref{sec:singleparts}). Finally, if increasing $l$ is not possible in these ways, as samples with large $\rho$ or $\varphi$ are to be investigated, the sample could be made very thin to minimize $L$ \cite{Born2014,Medebach2007}. Many colloidal samples are in this regime of single-scattering, as the particles are at least partially index-matched by the suspending liquid and the size parameter of the particles are small enough.
	
	\item[\emph{$l<L$}:] The mean free path is smaller than the sample size and the light is scattered multiple times before exiting the sample. The sample takes a turbid or milky-white appearance depending on the degree of multiple scattering and of absorption. The first-order approximation is not valid anymore, as the particles are excited by the incoming wave as well as scattered waves. A general analytical theory for light propagation in the case of multiple scattering becomes very involved and is subject of ongoing research \cite{Kristensson2015,Ishimaru1999}, and direct determination of the structure factor of a particle ensemble seems not possible anymore. 
		
	\item[\emph{$l\ll L$}:]  In the extreme of $l\ll L$ the light propagation reaches the limit of light diffusion, in which the incoming wave is fully scrambled and light propagation follows random walk statistics \cite{Ishimaru1999}. The statistics of this random walk are parametrized by a diffusion coefficient for light $D = c l^{\ast}/3$, where $c$ is the speed of light and $l^{\ast}$ is the randomization length, the distance the light has to travel before it is randomized again. The light transport in the limit of light diffusion is important for the approach to scattering-based analysis of granular media, and is therefore described in detail:
	
	\begin{description}
		\item[\emph{$l>l_{c}$}:]  A local first-order approximation could be made if $l$ is at least larger than any spatial correlation length $l_{c}$ of the sample (this is the situation sketched in Fig.~\ref{fig:bulk}). The sample could be assumed to consist of many independent single-scattering samples \cite{Weitz1993,Fraden1990,Rojas2004}. A diffusion-like light propagation and the applicability of a local first-order approximation allows for diffuse-transmission spectroscopy (DTS) \cite{Kaplan1994}, which is discussed in Sec.~\ref{sec:diffusion} in more detail. The method allows for verifying predicted structure factors \cite{Kaplan1994}. Samples that regularly fulfill the conditions for DTS are large colloidal samples, such that partial index-matching and particle size allow for large a $l$, but still strong multiple scattering occurs within the sample.
		
		\item[\emph{$l<l_{c}$}:] The light is scattered multiple times even within distances where particle positions are correlated. The first-order approximation cannot be applied even locally to a cluster of correlated particles anymore and the scattering patterns are considerably altered. An example for scattering in such a regime are Kikuchi patterns, which are created by multiple scattering within a single-crystalline region \cite{Nisbet2015}.
		
		\item[\emph{$l<\lambda$ }:] In this regime an extreme situation of wave propagation is predicted. Waves may become fully localized when the direction of the wave is randomized on distances shorter than the wavelength. The scattering objects have to be smaller than the wavelength but still have to have a very high scattering efficiency to observe this behavior, a combination which is difficult to match. Presently this behavior has not been observed and seems unlikely to occur in particle packings \cite{Skipetrov2016,Sperling2016}. 
	\end{description}
\end{description}

Further constraints are set by ratio of the wavelength and structural distances in the sample:

\begin{description}
	\item[\emph{$\lambda\ll r$}:] The wavelength is much shorter than the distance $r$ between obstacles or surfaces in the sample. In this regime, called far-field scattering, the scattered wave can be approximated as an outgoing spherical wave, which is prerequisite to the analysis used in the first-order approximation \cite{Born1999}.
	
	\item[\emph{$\lambda\geq r$}:] Electromagnetic interaction between objects happens by the evanescent, non-propagating near-field component of the scattered wave when they are in close proximity. Propagation properties have to be corrected for this near-field coupling, which may not be an established method presently \cite{Petrova2009,Schaefer2012,Rezvani2015}.
	
	\item[\emph{$\lambda\approx l_{c}$}:] The wavelength matches a correlation length in the sample and Bragg-scattering occurs. Bragg-like scattering also emerges in disordered materials with strongly correlated particle positions, and affect strongly the possibilities for wave propagation \cite{Rojas2004,Froufe2016}. The modulation of the density of possible wave states in correlated media is reminiscent of formation of a band structure.
	
	\item[\emph{$\lambda\gg l_{c}$}:] The sample effectively behaves like a homogeneous medium when the wavelength in the experiment exceeds any spatial correlation length in the sample \cite{Hapke1993}. The medium then has an effective refractive index which could be predicted by effective medium approximations. A widely used approximation to calculate an effective refractive index is the Maxwell-Garnett equation,
\begin{equation}
m_{eff}^{2}= m_{m}^{2}\frac{2\phi(m_{p}^{2}-m_{m}^{2})+m_{p}^{2}+2m_{m}^{2}}{2m_{m}^{2}+m_{p}^{2}+\phi(m_{m}^{2}-m_{p}^{2})}
\label{eq:MG}
\end{equation}	
where indexes $m$ and $p$ indicate host medium and particles, $\phi$ is the volume fraction occupied by particles, and absorption is neglected \cite{Kolokova2001}. Measuring the refractive index of a sample thus in principle allows retrieving the volume fraction of particles in the sample. More complex effective medium theories may have to be applied, which also take correlated particle position into account to calculate effective refractive indexes \cite{Tsang2000}. 
\end{description}

\subsection{The first-order approximation}
\label{sec:firstorder}

The \emph{first-order approximation} plays a prominent role in the established scattering-based methods as it fundamentally relates structure and scattering properties. It is based on two assumptions \cite{Born1999}. One is that the wave scattered by the sample is observed only far from the sample, so that the scattered wave can be approximated as an outgoing spherical wave. The other assumption is that within the sample the total electromagnetic field can be approximated by the incoming wave. The scattering amplitude function turns into the Fourier transform of the scattering potential of the sample within the validity of these two assumptions. The scattering potential describes the spatial distribution of the refractive index or alternatively of the dielectric coefficients within the sample. 

These approximations are also termed \emph{Rayleigh-} and \emph{Rayleigh-Gans-Debye-scattering} when the sample consists of individual particles \cite{Hulst1981,Bohren1983}. Each dipole of a particle is only excited by the incoming wave and the scattered field is observed far away from the sample. The approximations are also called \emph{single-scattering-} or \emph{Born approximation} with samples consisting of ensembles of many particles, as all particles only scatter the incoming light and not the light scattered from other particles in the medium \cite{Born1999}.

Measuring the scattering amplitude of a sample in all spatial directions, determined by the azimuthal angle $\theta$ (the polar angles will be averaged for isotropic media) or the momentum transfer vector $\vec{q} = 4\pi/\lambda\cdot\sin(\theta/2)$, allows calculating a low-pass filtered approximation of the scattering potential of the sample \cite{Born1999}. This is especially interesting for samples which consist of identical objects whose scattering potential can be described as the convolution of the scattering potential of a single object and a set of Delta-functions. The Fourier-transform of the scattering process turns the convolution into a product of the form factor of the objects $F(|\vec{q}|)$ and the structure factor of the sample $S(|\vec{q}|)$ \cite{Feigin1987}. 
\begin{equation}
	I_{s}(\vec{q}) \propto F(|\vec{q}|) \cdot S(|\vec{q}|)
\label{eq:firststorder}
\end{equation}
Measuring the scattering amplitude of dilute samples allows determining $F(|\vec{q}|)$ of the particles. Measuring the scattering amplitude of the dense sample and division by $F(|\vec{q}|)$ consequently allows isolating the structure factor of the particle packing \cite{Feigin1987}.

The structure factor is the momentum transfer- or reciprocal space Fourier-transform of the radial distribution function \cite{Feigin1987}. The structure factor of disordered hard-sphere packings exhibits a pronounced peak at a momentum transfer of $|\vec{q}|= 2\pi/d$ \cite{Hansen2005}, where $d=2a$ is the particle diameter. This indicates a pronounced correlation length $l_{c}$ within the packing with a length of $d$. The hard-sphere structure factor exhibits some trailing oscillations, whose amplitude decay as $|\vec{q}|^{-2}$, and are barely visible beyond $|\vec{q}|\approx 10\pi/d$ \cite{Hansen2005}.

\subsection{The diffusion approximation}
\label{sec:diffusion}

The diffusion approximation gives a solution to the radiative transfer equation in the limit of strong multiple scattering \cite{Ishimaru1999}. A relation among the packing structure as described by $S(|\vec{q}|)$ and the transmitted intensity can also be established in this regime. The central quantity of diffusion-like radiation transport, the random walk step length or \emph{randomization length} $l^{\ast}$, is determined by the scattering anisotropy of the individual scattering events. It takes many scattering events to randomize the light propagation direction if scattering is very anisotropic and confined to the forward direction (compare Sec.~\ref{sec:singleparts}), and $l^{\ast}\gg l$. If the light is isotropically scattered at each scattering event, $l^{\ast}$ becomes equal to $l$.

A relation among the randomization length and the packing structure in the sample can be obtained in situations in which a local first-order approximation is valid \cite{Weitz1993}. This implies that the light scattered by a cluster with correlated positions is only scattered again in the far-field of this cluster ($l>>l_{c}$), and that within a cluster with correlated positions the single-scattering approximation holds. The sample could be considered an ensemble of independent clusters, whose scattering amplitude can be described by $F(|\vec{q}|) \cdot S(|\vec{q}|)$ \cite{Kaplan1994}.

The relation among $l^{\ast}$ and $l$ is given by the scattering anisotropy or the average cosine of the azimuthal scattering angle of the scattering events within the validity of the local first-order approximation \cite{Fraden1990}:
\begin{eqnarray}
	l^{\ast} &=& l / \left<(1-\cos(\theta))\right> \\ &=& \frac{1}{\varphi}\cdot \frac{1}{\int_{0}^{4\pi/\lambda}{F(|\vec{q}|)S(|\vec{q})|(1-\cos(\theta))|\vec{q}|dq}}
\label{eq:lstar}
\end{eqnarray}
The spectral transmission $T(\lambda) = \frac{I_{s}}{I_{0}}$ through a sample with diffusion-like light transport is then, in the limit of negligible absorption, proportional to the ratio of $l^{\ast}$ and the sample thickness $L$ \cite{Kaplan1994}:
\begin{equation}
	T(\lambda) = \frac{I_{s}}{I_{0}}\propto\left(\frac{l^{\ast}(\lambda)}{L}\right) \frac{c_{1}}{1+c_{2}\left(\frac{l^{\ast}(\lambda)}{L}\right)},
\label{eq:T}
\end{equation}
where the coefficients $c_{1}$ and $c_{2}$ depend on the reflection coefficients from the sample boundaries and a geometrical factor for the light source. From Equations \ref{eq:lstar} and \ref{eq:T} it follows that spectral transmission measurements probe integrated properties of the scattering amplitude of the sample, what forms the basis of diffuse-transmission spectroscopy (DTS). A direct determination of $S(|\vec{q}|)$ as in the regime of single-scattering is not possible, but predicted structure factors can be validated \cite{Kaplan1994}. 

The upper integration bound in Eq.~\ref{eq:lstar} causes the spectral structure sensitivity of DTS. The structure factor of hard-sphere packings barely exhibits any features beyond $|\vec{q}|\approx 10\pi/d$ with the main peak at $|\vec{q}|= 2\pi/d$ \cite{Hansen2005}. There will be little to none spectral variation in transmission if $4\pi/\lambda$ is much larger than $10\pi/d$. The spectral transmission measurement consequently will be most sensitive to the variations in the structure factor if wavelengths from a fraction of $d$ up to a few times $d$ are used. A combination of visible light and granular media, in contrast, would lead to integration boundaries which are orders of magnitude larger than where the last features of $S(|\vec{q}|)$ are observable. The spectral transmission of visible light through granular media thus is basically independent of the packing structure within the sample, as can be observed by the common white appearance of distinct samples like snow, sugar, flour or salt. DTS will become sensitive to the structure of granular media again when electromagnetic radiation with $\lambda\approx d$ is applied, as is discussed in the following.

\subsection{Classification of granular media}
\label{sec:granmedia}

The scattering regimes and the approximations required for obtaining structural information were schematically defined in the preceding sections. It is instructive to now classify a typical granular medium according to above scheme. 

\begin{figure}
	\centering
		\includegraphics[width=0.5\textwidth]{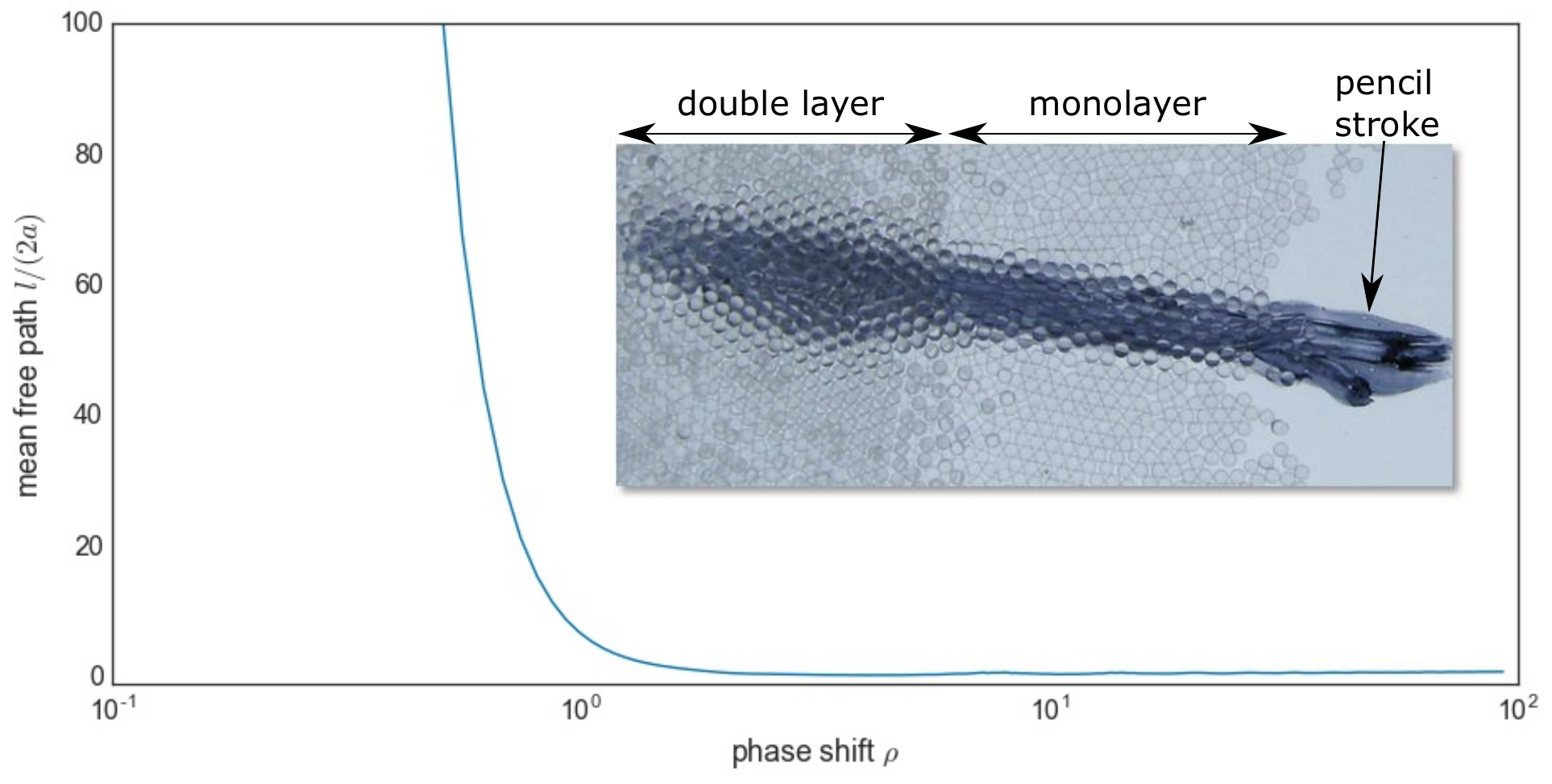}
	\caption{The mean free path $l$, normalized by the diameter of of the spheres $2a$, of a dense packing according to eq.~\ref{eq:l}. Assumed are a volume fraction $\Phi = \phi\cdot4/3\pi a^{3}=0.64$ and the single-particle scattering efficiency $Q_{s}$ displayed in Fig.~\ref{fig:Q}. The mean free path in a dense packing rapidly decays to a single particle diameter for particles larger than the wavelength. These extremely short $l$ are illustrated in the inset by a pencil stroke imaged through one and two layer of transparent particles with 500~$\mu$m diameter. The difference among the blurred stroke imaged through the monolayer and the double layer show a redirection of the light at each particle layer.}
	\label{fig:stroke}
\end{figure}

Classification requires the mean free path $l$ within the sample and the wavelength $\lambda$. The mean free path $l$ can be estimated for a granular packing using Eq.~\ref{eq:l}. Granular media are conventionally characterized by a close-packing of the particles, leading to high packing fractions $\phi$ around the close-packing limit of monosized spheres of $\approx$0.64. The number density $\varphi$ and the volume fraction $\phi$ are connected by $\varphi = \phi/(4/3\pi a^{3})$. The scattering efficiency is well approximated by being between 1 and 2 over broad ranges of wavelengths or particle sizes (unless the wavelength is much larger than the particles or index matching is used, see Sec.~\ref{sec:singleparts}). Using these numbers in Eq.~\ref{eq:l} gives values for $l$ of around a particle diameter. This is an upper bound due to the high assumed numbers for $\phi$ and $Q_{s}$, but this number still indicates that in the case of granular media the mean free path is very short up to the light changing direction at each particle surface (compare Fig.~\ref{fig:stroke}). This extremely short $l$ makes scattering from granular media different from the situation of colloidal, molecular or atomic media combined with X-ray scattering or visible light scattering. The small relative refractive indexes ($|m-1|$ in Eq.~\ref{eq:rho}) and the possibility to dilution regularly lead to large $l$, both compared to $l_{c}$ and to $L$.

The evaluation by the first-order approximation requires single-scattering ($l>L$) and far-field propagation ($\lambda\ll d$) (see Sec.~\ref{sec:firstorder}). These conditions can only be reached by a particle monolayer for dense granular media \cite{Born2014} or dilution to sparse particles \cite{Xu2002}. The scattering regime of the medium will be single-scattering nearly independent of the wavelength in these situation, while the wavelength adjusts whether RGD-, Mie- or Fraunhofer scattering theory has to be applied for the particles.

On the other hand, bulk granular media will regularly fulfill the prerequisite $l\ll L$ for diffusion-like wave transport. The wavelength used in the scattering experiment will have a strong influence whether $l<l_{c}$ or $l>l_{c}$ and whether $\lambda\ll r$, $\lambda\approx l_{c}$ or $\lambda> l_{c}$ will be reached, and thus whether DTS could be applied to the medium. 

\subsection{Diffuse-transmission THz spectroscopy}
\label{sec:diffuseTHz}

Typical grain sizes in experiments with granular media are a few tens of micrometer up to a few millimeter. These particle diameters correspond to wavelengths from the THz region of the electromagnetic spectrum. Spectroscopic THz transmission measurements will be sensitive to the local structure of granular samples, as $\lambda$ varies from fractions of $d$ to a few times $d$ (see Sec.~\ref{sec:diffusion}). The problem of scattering in bulk granular media has attracted interest soon after the more widespread introduction of THz spectrometer. Scattering effects in powders were complicating the evaluation of THz spectra in pharmaceutical or security applications \cite{Taday2004,Bandyopadhyay2007,Zurk2008}. Correction for the scattering effects is required to quantitatively extract absorption features of molecular substances \cite{Shen2008,Kaushik2012}. It was noted that THz extinction spectra on granular samples change when the samples start flowing, what indicated a structure sensitivity of the spectra \cite{May2009}. Transmission experiments with THz radiation were also used as scale model experiments for scattering from dense particulate media in the visible spectral range \cite{Pearce2004,Pearce2001}.

However, with the mean free path $l$ being roughly a particle diameter (see Sec.~\ref{sec:granmedia} and Fig.~\ref{fig:stroke}) and $\lambda$ being roughly a particle diameter, $l>l_{c}$ and $\lambda\ll r$ will not be reachable for densely packed media. The approximations made for the extablished DTS methodology thus will not be fulfilled for granular media in general. We suggest that transmission spectroscopy can still be used to obtain structural information on bulk granular media. No propagating wave is possible when $\lambda$ matches the Bragg-condition for backscattering ($\theta = \pi$), $|\vec{q}|\cdot d = 2\pi$ or $\lambda = 2d$. In this case only standing waves formed by the incident and reflected wave are possible \cite{Gerthsen2010}. Hard-sphere packings form isotropic Brillouin zones with radius $2\pi /d$ \cite{Froufe2016}, as can be seen from the peak of the structure factor. Propagation of waves with $\lambda = 2d$ thus will be suppressed isotropically, and a photonic bandgap might be formed \cite{Froufe2016,Takagi2004,Takagi2010}. Measuring the wavelength of the suppressed transmission reveals the position of the structure factor peak. The degree of extinction of the propagating wave is dependent of the development of the structure factor peak and thus the degree of correlation in the sample \cite{Rojas2004,Froufe2016}


\section{Scattering of THz radiation: Materials and Methods}
\label{sec:THzscatt}

Two approaches to probe the scattering by granular particles and granular particle packings were highlighted in the previous section. First, angle-resolved measurements can only be performed and analyzed using the first-order approximation with particle monolayer samples or extremely dilute samples. Such measurements can be performed using specialized small-angle laser light setups with visible light \cite{Xu2002}, but also in the THz range \cite{Born2014}. Second, spectroscopic measurements on Bragg-scattering resonances can be performed on bulk samples using THz radiation. Common prerequisites for such measurements are low absorption and large illuminated areas.

The Terahertz spectral region covers the range from 0.1~THz to 10~THz, or 3~mm to 30~$\mu$m, respectively \cite{Brundermann2011}. Some properties of THz are similar to the infrared and visible region, and similar optical components like mirrors and lenses can be used. On the other hand, properties of the microwave frequency region are pertinent, and components like antennas and waveguides are used. Books that provide broader and deeper information on THz methods and materials are available \cite{Brundermann2011,Peiponen2013,Naftaly2015}.

\subsection{Materials}
One important aspect that is shared among the THz region and microwaves is dielectric heating. Therefor absorption losses are frequently high in the THz region \cite{Brundermann2011}. Especially water and water vapor are strong absorbers, which limits free-space propagation of THz radiation in experimental setups.

 Useful materials for optical components and also particle packing experiments are unpolar polymers. In table~\ref{tab:materials} some common materials for experiments with THz radiation are listed. It is apparent, that even the optimal materials for experiments with THz radiation, like PTFE and PE, have much higher absorption coefficients than materials for optical components in the visible range have. 
\begin{table}%
\begin{tabular}{lccr}
\hline\hline
Material & $m'$ & $\alpha$~[cm$^{-1}$] \\
\hline
teflon (PTFE) & 1.44 & 2\ldots 3 \\
paraffin & 1.49 & 0.1\ldots 6  \\
polypropylen (PP) & 1.5 & 0.1\ldots 2  \\
polyethylene (PE) & 1.53 & 0.06\ldots 2 \\
polystyrene (PS) & 1.59 & 0.28\ldots 2 \\
plexiglass & 1.61 & 0.59\ldots 15  \\
quartz & 2.4 & 0.05\ldots 5  \\
window glass & 2.58 & 1.95\ldots 20  \\
sapphire & 4 & 0.1\ldots 20 \\
\hline
visible range: & & \\
window glass  & 1.47 & 0.000001  \\
\hline\hline
\end{tabular}
\caption{Overview over optical properties of common materials for THz experiments. Approximative values for the real part of the refractive index $m'$ and the absorption coefficient of the material $\alpha$ are given. These values are estimates from various sources \cite{Brundermann2011,Naftaly2015,Piesiewicz2007,Folks2007,Cunningham2011} and for a spectral range from 0.5~THz to 4~THz.}
\label{tab:materials}
\end{table}

\subsection{Methods}

THz radiation allows for setup design in close analogy to optical setups in the visible and infrared range. Optical components like lenses and mirrors can be obtained off-the-shelf\cite{Brundermann2011,Naftaly2015,Peiponen2013}. 

\paragraph{Angle-resolved measurements:} As discussed in Sec. \ref{sec:granmedia} and \ref{sec:firstorder}, conventional scattering experiments, where scattered intensity is measured against the scattering angle, will play a secondary role in characterization of granular packings. Such conventional scattering experiments require a collimated incoming monochromatic beam from some light source, a sample holder with sample and a detector on a goniometer to measure angle-resolved scattered intensities. 

Light sources which are conveniently used for scattering experiments are lasers, which produce monochromatic collimated beams. Laser for THz radiation are available with various pump principles and gain media. The most convenient THz lasers are probably quantum cascade lasers (QCL) \cite{Richter2010}. A beam expansion might be required when particle sizes reach the beam diameter, in order to average over meaningful particle numbers.

Sample cells are preferably made out of unpolar polymers like PE or PS, which have a nearly constant refractive index over most of the THz spectral range and moderate absorption (see tab.~\ref{tab:materials}). Thin PE-foil, which is available in kitchen stores, is a pretty good window material or sample holder. Crystalline quartz windows could also be used if higher thermal, chemical or mechanical robustness is required \cite{Brundermann2011}. These benefits come at the cost of a higher refractive index and thus higher reflection losses at the windows.

Golay cells are a workhorse for detection of THz radiation \cite{Brundermann2011}. They can be operated at room temperature, but are pretty slow. Semiconducting bolometers or photodectors are fast, but require cooling by liquid nitrogen or even liquid helium for highest sensitivity \cite{Brundermann2011}. The detector should be equipped with some collection and collimation optics when the source is equipped with a beam expander.
	
\paragraph{Spectroscopic measurements:} Spectroscopic measurement of the light transport properties seems the more promising approach for characterization of particles and particle packings which are dominated by multiple scattering. THz spectrometer are available off-the-shelf, thus one does not need to worry about source, detector and setup. The most common setup types for spectroscopy with THz radiation are Fourier-Transform (FT) spectrometer and THz-Time Domain Spectrometer (THz-TDS) \cite{Brundermann2011}. 

FT-THz spectrometer are compatible with FT-IR spectrometer. Only the semitransparent mirror and the detector are optimized for THz radiation, and the setup usually can be evacuated to minimize atmospheric absorption. FT-THz spectrometer provide excellent resolution down to 0.2~GHz, wide spectral range of 11 THz and short measurement time below a minute. The setups also usually provide modularity, so that source and detector in principle could be used in other experiments. FT-THz spectrometer can be combined with synchrotron radiation as light source. The spectrometer has then a pulsed, coherent light source, with superior intensity especially in the range of long-wave THz radiation \cite{Holldack2016}.

THz-TD-spectrometer use a pulsed source to probe the THz spectral range. They usually have resolutions of 1.2~GHz, a dynamic range of 75~dB around a frequency of 1~THz (which rapidly reduces for shorter and longer wavelengths), spectral ranges of 4~THz, and measurement times can be shorter than a minute. However, they have the benefits of being less expensive than FT-THz spectrometer, of a flexible measurement geometry that can easily be changed from reflection to transmission, of compactness which facilitates integration into other experiments, and of providing phase information, which means that the effective refractive index of the sample could directly be determined \cite{Brundermann2011}.

Decisive parameters for transport measurements on scattering samples are mainly high intensity, a high dynamic range, and broad spectral range. Strong intensity losses will happen due to scattering in the bulk samples and the non-neglectable absorption in THz range and require high intensities and high dynamic ranges. The spectral range of the setup directly determines the particle sizes and correlation lengths that can be investigated, thus a broad spectral range is highly desirable. High resolution, however, might be interesting for characterization of particles, as a fine ripple structure in the scattering spectra give information on size distributions \cite{Born2015}. Resolution is usually not an issue for characterization of packings, as $F(|\vec{q}|)$ and $S(|\vec{q}|)$ do not show abrupt changes or steep gradients (compare Sec.~\ref{sec:diffusion}).

\subsection{Optimization of THz spectroscopy}
\label{sec:optimization}
THz has potential to become a supplemental method to imaging methods for granluar media. Spectroscopy allows obtaining the transmitted intensity $I_{s}(\lambda)$. The transmission is defined as $T(\lambda) = I_{s}(\lambda)/I_{0}(\lambda)$, the ratio of measured intensity with sample to measured intensity without sample. Only few adjustments need to be made to conventional optical paths in order to optimize THz spectroscopy for granular media:

\begin{figure}
	\centering
		\includegraphics[width=0.5\textwidth]{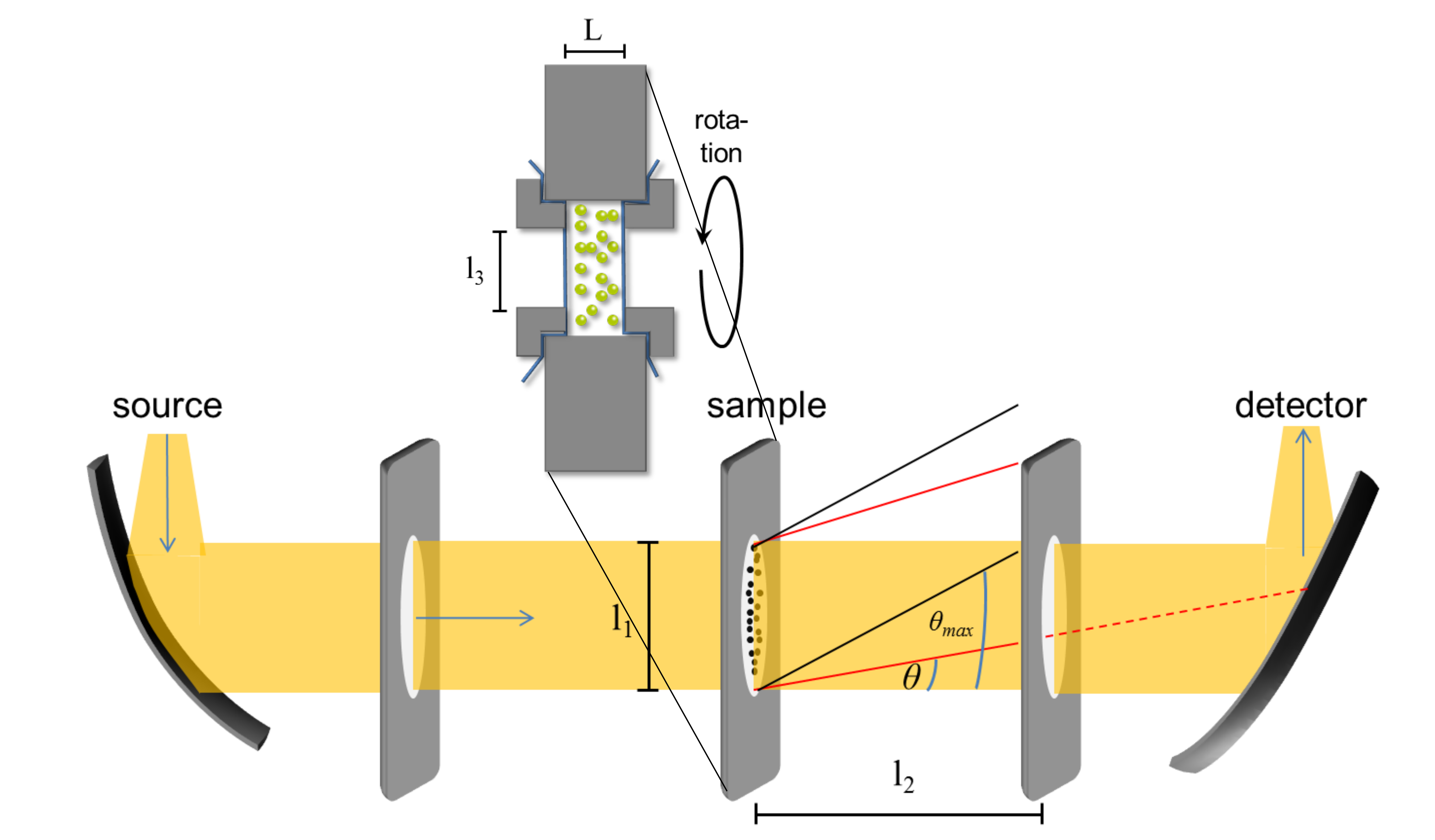}
	\caption{Scheme of light propagation in the FT-THZ spectrometer used in the experiments. An iris is used to block scattered light down to a minimal angle, below which it will reach the detector. The beam diameter $l_{1}$ is 10~mm, the distance to the collimating iris $l_{2}$ is 70~mm, allowing light scattered up to $\theta_{max} = 4^{\circ}$ to reach the detector. The sample cell consisted of a feedthrough which was sealed by thin PE foil hold in position in by two o-rings. The diameter $l_{3}$ of the feedthrough is 20~mm, the thickness $L$ in beam direction is 10~mm. The sample was rotated by a stepper motor to improve the measurements, see sec.~\ref{sec:optimization}.} 
	\label{fig:setup}
\end{figure}
\paragraph{Collimated detection vs. integrating sphere:} A typical optical path is sketched in Fig.~\ref{fig:setup}. Divergent radiation from a source is somehow collimated, passes through the sample and is focused onto a detector. Usually an iris or some other optical element limits the angle $\theta_{max}$ up to which scattered radiation is accepted by the detector. This is important for characterization of individual particles by measuring their extinction spectroscopy \cite{Born2015}. A large $\theta_{max}$ will lead to incorrect results in this situation (see Fig.~\ref{fig:Q}).

The situation is different for measurements in the limit of light diffusion. Only scattered light will be transported through the sample. An integrating sphere to measure all the diffusively transmitted light is required to quantitatively measure $T(\lambda)$ and $l^{*}(\lambda)$ \cite{Kaplan1994}. However, assuming a slab like geometry with much larger lateral dimensions than in beam direction, most of the light will leave the sample in a central region \cite{Leutz1996}, and measuring the transmission with a collecting lens as in THz-TD- or a collecting mirror as in FT-THz-spectrometer gives an reliable estimate of the spectral transmission properties.

\begin{figure}
	\centering
		\includegraphics[width=0.5\textwidth]{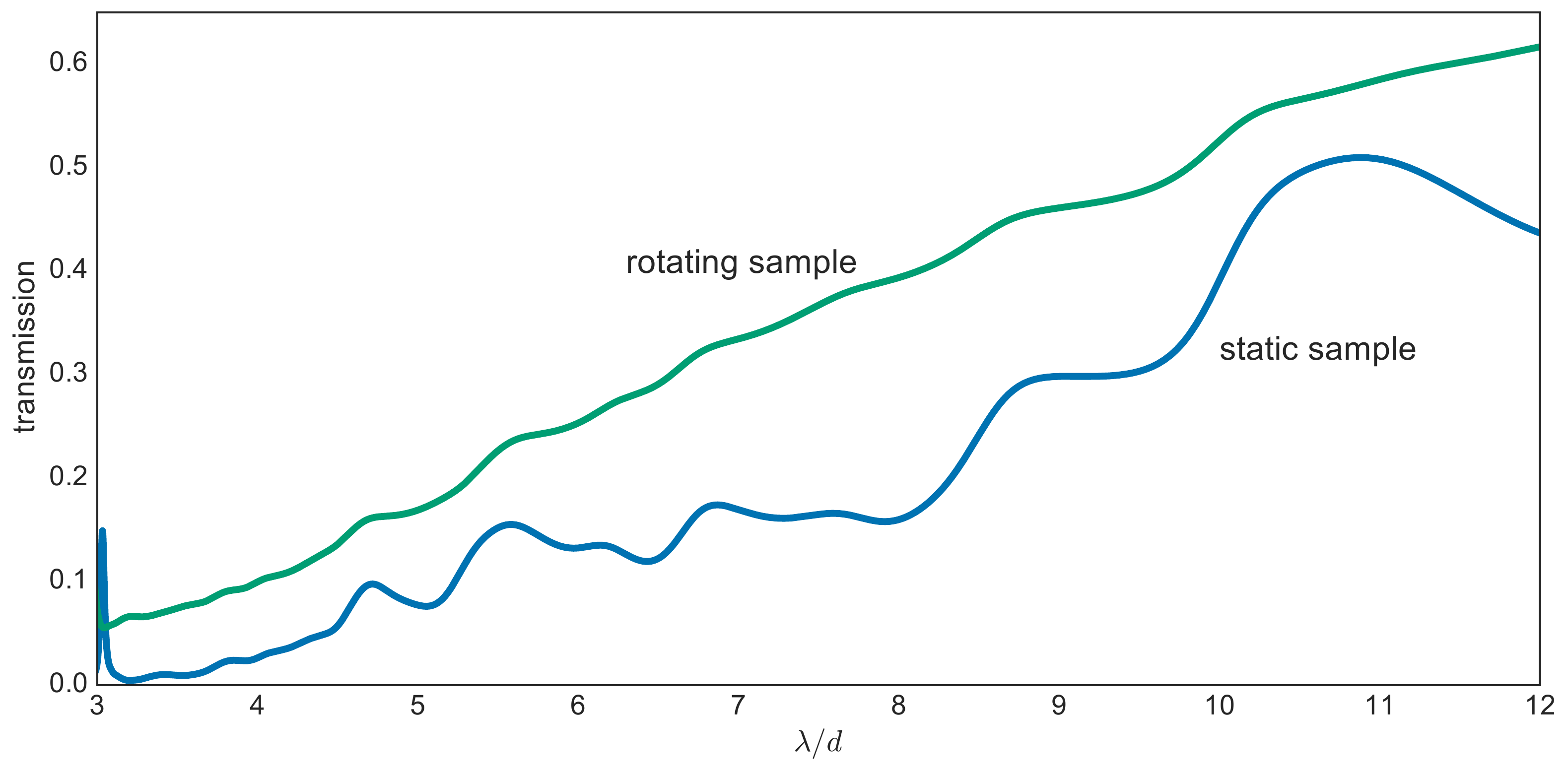}
	\caption{Comparison of transmission spectra measured with coherent THz radiation and with and without rotating the sample. Rotation smears out the irregular speckle pattern and allows for measuring smooth transmission spectra. The curves are offset for clarity.}
	\label{fig:rotation}
\end{figure}
\paragraph{Rotation for configuration averaging and speckle washing:} Pulsed light sources like synchrotron sources or time domain spectrometer generate coherent THz radiation \cite{Abo2003}. This will lead to an irregular intensity distribution, a speckle pattern, after transmission through the sample. This spatial irregularity also may become manifest in the measured spectra when no integrating sphere is used. Rotation of the sample smears out the speckles and flattens the spectra (see Fig.~\ref{fig:rotation}). Rotation of the sample during the measurement also increases the number of configurations from which spectra are measured and increase the reproducibility of the measured spectra.

\begin{figure}
	\centering
		\includegraphics[width=0.5\textwidth]{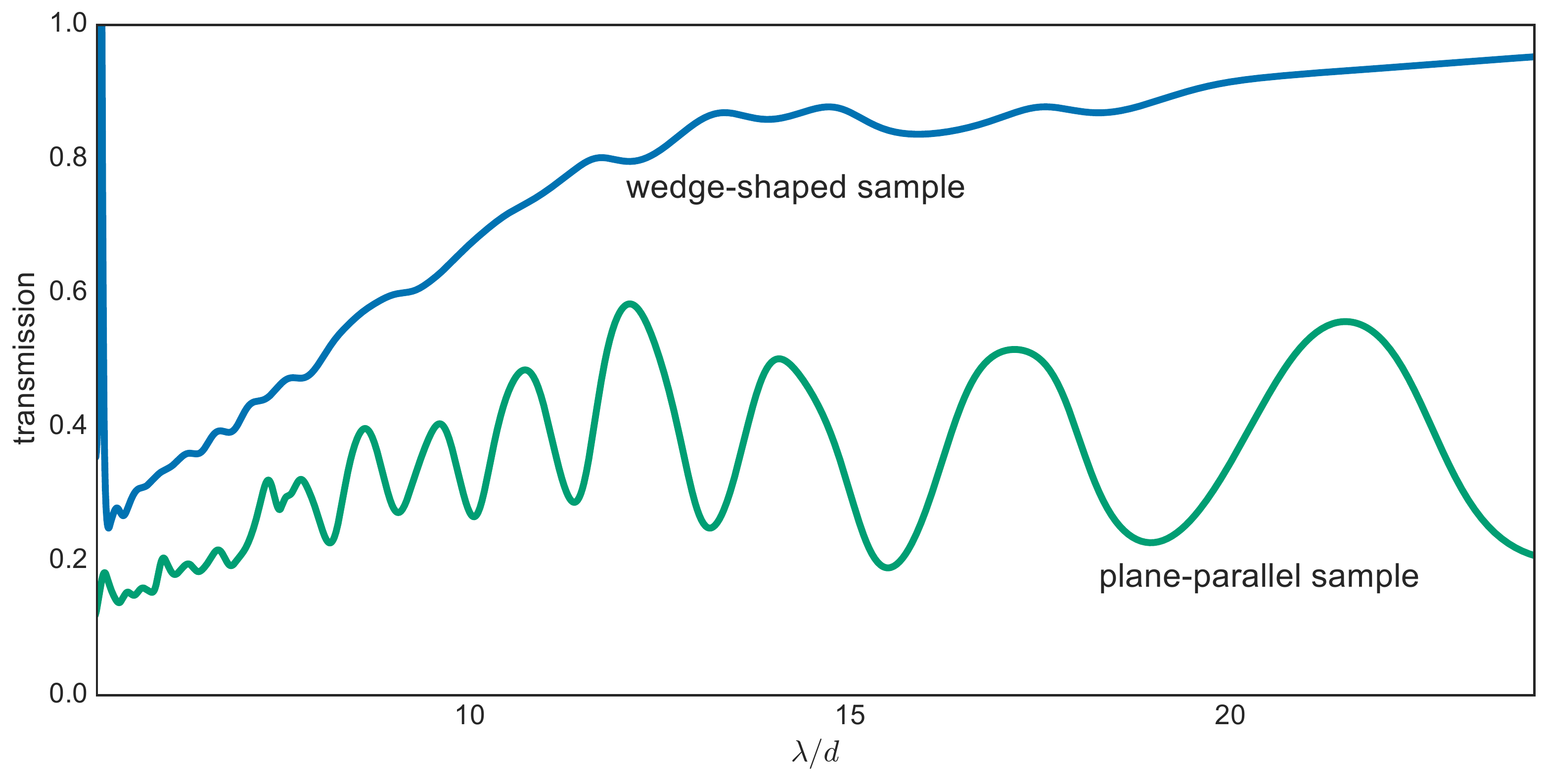}
	\caption{Comparison of transmission spectra obtained with a wedge-shaped sample cell and with flat parallel windows. The cell with flat parallel windows forms an etalon for the THz radiation, leading to oscillatory transmission values. The oscillations are effectively supressed with the wedge-shaped sample cell. The curves are offset for clarity.}
	\label{fig:etalon}
\end{figure}
\paragraph{Fabry-Perot etalons:} A conventional sample cell with flat parallel entrance and exit windows forms a nice etalon for the large wavelengths from the THz spectrum. Oscillatory fringes from the sample cell etalon will be superimposed to the spectral extinction from the sample. These etalon effects could be either corrected numerically \cite{Withaya2006}, or the sample cell windows are very gently misaligned to a wedge-shaped sample cell (see Fig.~\ref{fig:etalon}). This will give a moderate increase in reflection losses, but efficiently suppresses etalon fringes.


\section{Experiments}
\label{sec:experiments}

The discussion in Sec.~\ref{sec:basics} leads to the conclusion that gaining structural information on bulk granular media with scattering-based methods is not straight forward. The assumptions made in established methods to receive structural information are not fulfilled in granular media. Scattering still is expected to leave an imprint on transmission spectra. The Bragg-condition for backscattering reads as $|\vec{q}|d=2\pi$ or $2d/\lambda = 1$ for a hard-sphere packing with a structure factor peak at $|\vec{q}|=2\pi/d$. Around this condition a suppression of light transmission can be expected, whose degree depends on the development of the structure factor peak (Sec.~\ref{sec:diffuseTHz}). We demonstrate this effect and its potential for characterization of granular packings.

\subsection{Experimental implementation} 
\label{sec:implementation}
Transmission spectra were measured using a Bruker IFS 125 HR spectrometer at the THz-beamline of the BessyII storage ring (Berlin, Germany) \cite{Holldack2016,Holldack2007}. Synchrotron radiation in the \emph{low-$alpha$}-mode of the storage ring was used as light source. The instrument was evacuated during the measurements to minimize absorption by air and humidity. Spectra were usually taken at moderate spectral resolution of 30~GHz. The acquisition times were a few seconds per scan and roughly 3 min for spectra consisting of 100 averaged scans. A 6~$\mu$m multilayer beamsplitter was used in combination with a 4.2 K Bolometer (IR-Labs) as detector.

The granular media samples where 500~$\mu$m polystyrene (PS) spheres or mixtures of 500~$\mu$m and 80~$\mu$m PS spheres. The sample with free particles consisted of a thin polyethylene (PE) foil perpendicular to the THz beam, to which individual PS particles where electrostatically adhered. The bulk samples where produced by pouring particles into a cylindrical plastic container. The two flat surfaces of the cylinder where made of PE foil for transmission measurements, the cylinder height in beam direction $L$ was 1~cm. The windows were slightly misaligned for suppression of etalon effects. The particles formed a dense, disordered packing just after pouring into the cylinder. After tapping the sample container a sufficient amount of time the particles arranged into a few crystalline regions. The particles were also treated with an isopropanol-paraffin solution, which made the particles very sticky. These sticky particles formed a very loosely packed disordered packing upon pouring into the container.

The wavelength is rescaled by the particle diameter for displaying the results, which emphasizes the structural length scales that affect the spectra. The transmission value of the free particles is scaled by the ratio of the area covered by particles to the total beam diameter to correct for low density of particles.

We expect scattering features at wavelengths comparable to $2d$. This condition has to be met inside of the sample, while the wavelength is measured outside of the sample. The measured wavelength of any scattering feature has to be scaled by the refractive index of the sample to obtain probed structural length scales in the sample. Obtaining an effective refractive index of a granular sample is not trivial. The spectral range of interest here ($\lambda\approx 2d$) is beyond the limit of the validity of effective medium theories (compare Sec.~\ref{sec:media} and Eq.~\ref{eq:MG}). Still, a general behavior like suggested by Eq.~\ref{eq:MG} has to be expected, a refractive index of the sample changing from the vacuum value at low densities up to a refractive index close to bulk polystyrol at high packing densities.

\subsection{Experimental results} 
\begin{figure*}
	\centering
		\includegraphics[width=0.9\textwidth]{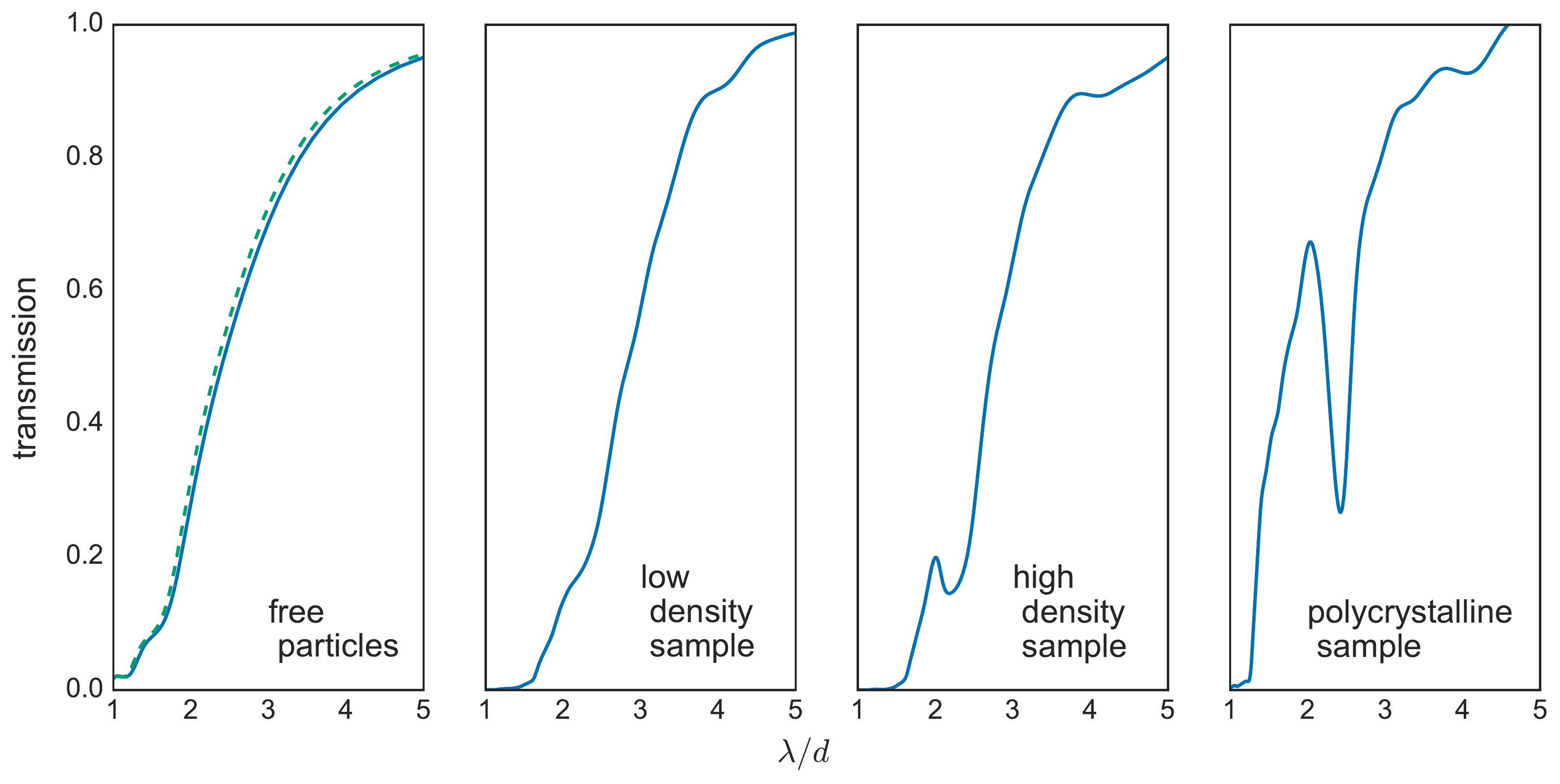}
		\caption{Demonstration of the structure sensitivity of THz transmission spectroscopy. Transmission is measured for individual particles, a low-density packing formed by sticky particles, a dense disordered packing formed by hard-spheres, and a polycrystalline packing of these spheres. Transmission becomes increasingly suppressed by Bragg-scattering at $\lambda\approx2d$ with increasing packing density. The position of the scattering feature shifts away from $\lambda=2d$ due to the increase of the effective refractive index of the particle packing with packing density. The dashed line in the left panel is the expected transmission of free particles with a fitted number density.}
	\label{fig:structure}
\end{figure*}

Figure~\ref{fig:structure} shows transmission measurements of granular sample with various degrees of order. The free particles show a behavior as expected from their cross-sections (Fig.~\ref{fig:Q}). The cross-sections become very small for wavelengths larger than the particles, and measured transmissions are high. Transmission values on the other hand become very small when the cross-sections increase for wavelengths equal or smaller than the particles. 

The transmission spectra of bulk packings exhibit the same general appearance, transmission close to 1 for wavelengths much larger than the particles, and ceasing transmission for wavelengths equal and shorter than the particles. A general increase in transmission with increase in packing density can be observed when comparing the bulk media. This might be an effect of the increasing refractive index of the sample and thus reduction of the scattering efficiency. The increasing correlation of the particle positions with packing density induces Bragg-like scattering, which increasingly lowers transmission at the scattering resonance condition of $\lambda = 2d$. The measured position of the scattering feature $\lambda\approx 2.5d$ and the predicted $\lambda\approx 2d$ differs, most likely due to the different wavelength within the sample. Using Eq.~\ref{eq:MG} to get a tentative refractive index of a PS sample with $\phi\approx 0.64$ gives a refractive index of 1.3. This cannot be exact as the approximations made for eq~\ref{eq:MG} are not fulfilled, but it illustrates that the shift in refractive index can be high enough enough to explain the difference. $\phi$ is even higher the crystalline arrangement, which consequently increases $m_{eff}$ and the measured peak position further. 

\begin{figure}
	\centering
		\includegraphics[width=0.45\textwidth]{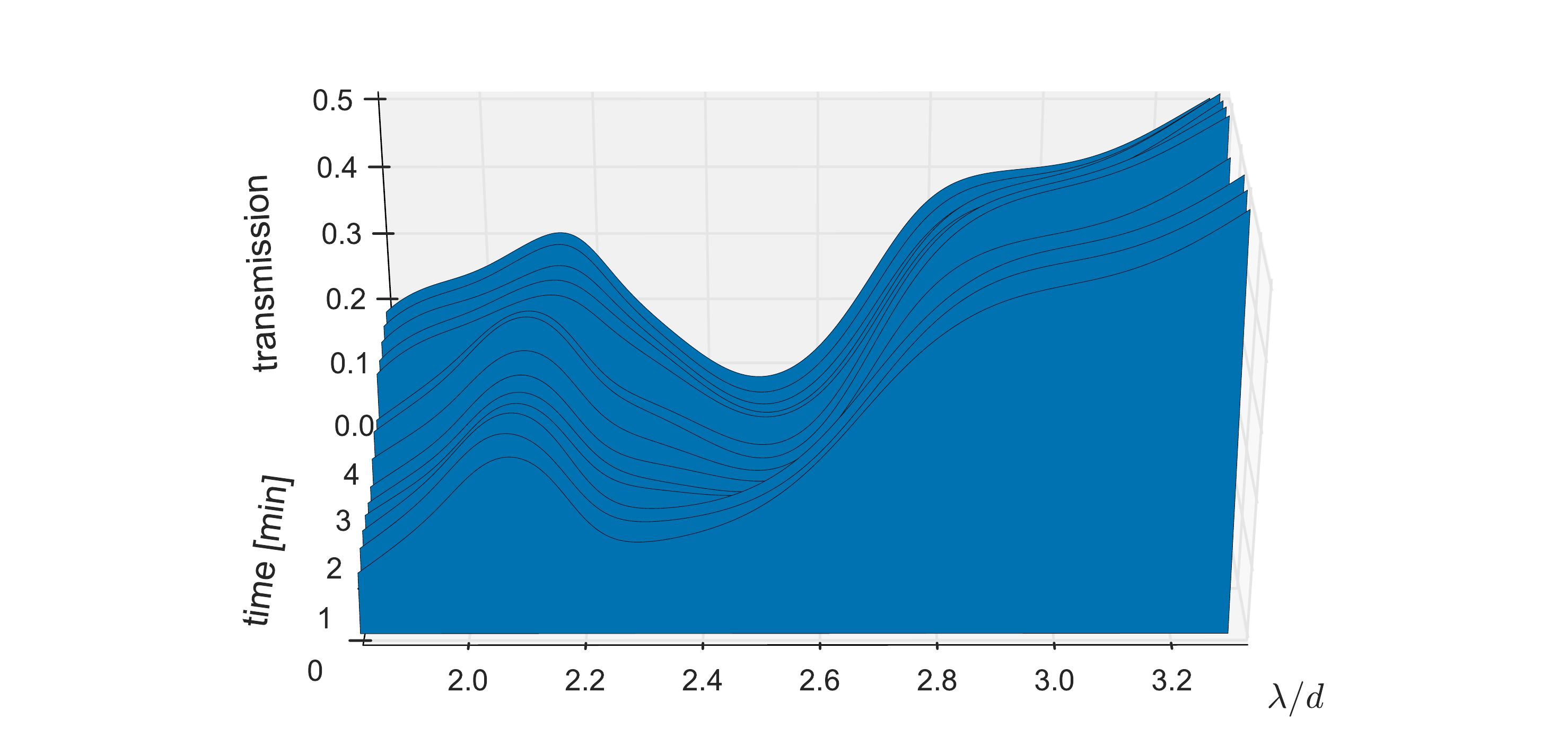}
		\caption{Evolution of the transmission spectrum of a vibrated packing of monosized spheres with time. The initial position of the scattering feature ($\lambda\approx 2.25d$) indicates a dense disordered packing, which evolves into a polycrystalline packing with a strong scattering feature at $\lambda\approx 2.6d$ within a few minutes.}
	\label{fig:time}
\end{figure}
\begin{figure}
	\centering
		\includegraphics[width=0.45\textwidth]{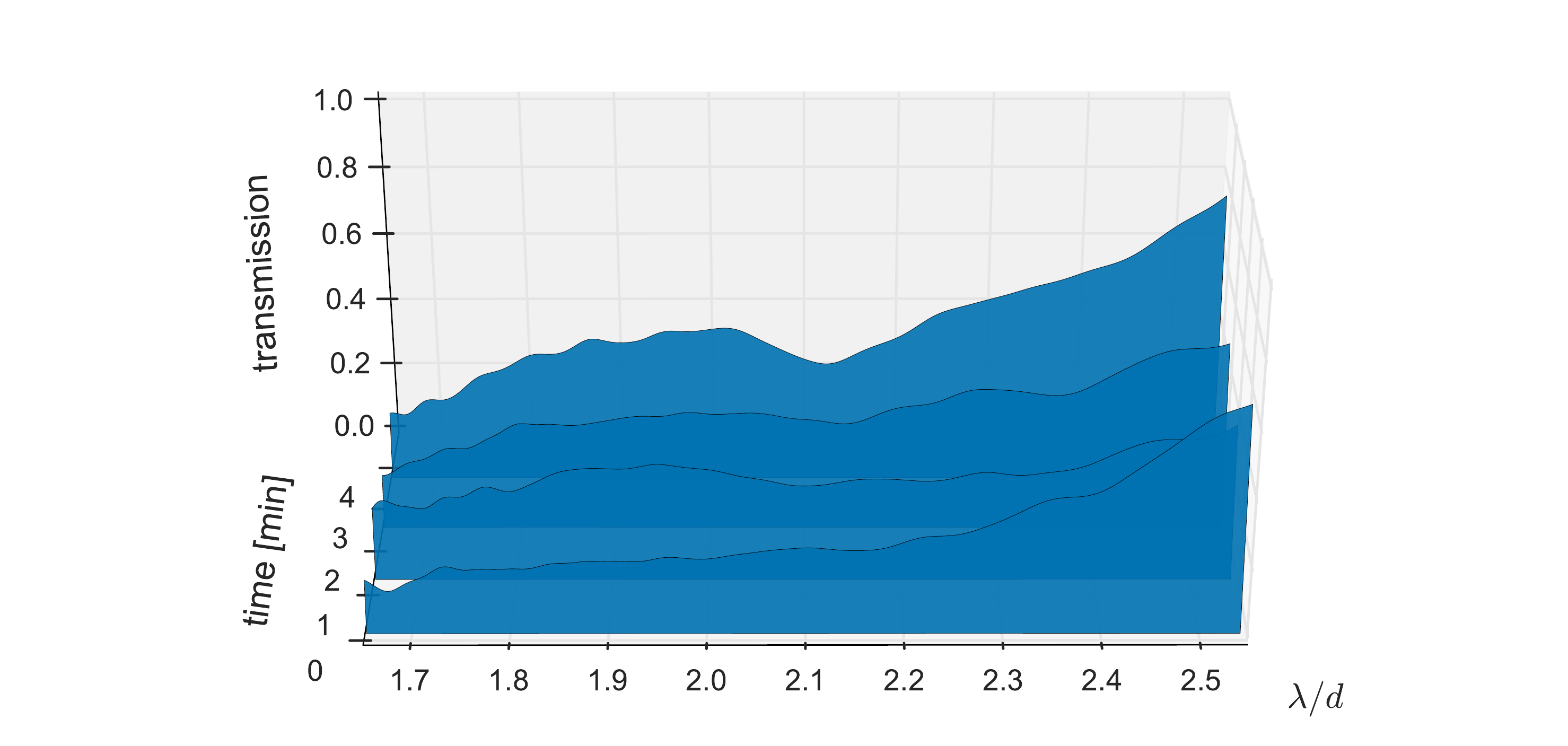}
		\caption{Evolution of the transmission spectrum of a vibrated binary mixture of spheres with size ratio 5:0.8. The spectra are focused on the spectral region for Bragg-scattering within a packing of the small particles. Emergence of a scattering feature at $\lambda\approx 2.15d$ after a few minutes of vibration indicates segregation of the two particle sizes and formation of regions with densely packed small particles.}
	\label{fig:demix}
\end{figure}

Summarizing, the transmission spectrum is characteristic of the packing density and the correlation length in the sample. This could be used to track structural changes in samples. Two examples are given in Figures \ref{fig:time} and \ref{fig:demix}. A small vibration motor is attached to the sample cell for the measurements in Fig.~\ref{fig:time}. The sample cell is again filled with monodisperse 500~$\mu$m PS spheres. The vibration motor is started with starting the measurements. The vibration induces some mobility of the particles and the whole packing can evolve from a dense, disordered packing into a crystalline packing. This evolution can be observed in Fig.~\ref{fig:time}. The scattering feature is at the beginning of the measurements at the position for a dense, disordered packing as in Fig.~\ref{fig:structure}. This feature transforms into the feature of the crystalline packing within a few seconds. It may be subject of further work to extract crystallization kinetics from such measurements.

The sample cell was filled with a binary mixture of 500~$\mu$m PS spheres and 80~$\mu$m PS spheres with a 1:1 volume ratio in another experiment (Fig.~\ref{fig:demix}). The wavelength is rescaled by the diameter of the smaller particles. Again, the measurements are started together with the vibration motor. At the beginning the spectrum is rather featureless, indicating that no dominant correlation length is present in the sample. A scattering feature in transmission evolves after a few minutes of vibration. This indicates that the sample has started to segregate and clusters with predominant small-small particle contacts emerge. Interestingly, segregation in this binary mixture happens on longer time scales than crystallization of the monosized spheres (Fig.~\ref{fig:time}). This might indicate higher kinetic barriers for segregation than for crystallization. Future work may use transmission spectra to track segregation processes and other dynamics in fluidized granular media in more detail.


\section{Conclusion and outlook}

Dense granular media exhibit very short mean free paths for most spectral regions. Particles are densely packed up to mechanical contact. These two issues prevent application of established scattering methodology to granular media. Spectroscopic THz transmission measurements are a promising approach to probe structural properties of granular packings with a scattering-based method. The low instrumental effort for such experiments will be minimized with spreading of benchtop solutions like Thz-TDS. We highlight this potential by THz transmission measurements on static packings with different degrees of spatial correlations. Different packing structures can be easily discriminated through their transmission spectra. We also demonstrate the possibility for monitoring transient structural states in agitated granular media with good time resolution. An important advancement required for further application of the presented approach is developing a reliable model for the effective refractive index over large ranges of $\lambda/d$-values. Finally, it may be noted that THz transmission spectroscopy offers the possibility to investigate photonic properties of disordered media on length scales where real space structures can be monitored and manipulated more easily than on length scales of visible light. 


\begin{acknowledgments}

P. B. thanks Matthias Sperl and Andreas Meyer for their continued support of the project, Jan Haeberle and Sebastian Pitikaris for help during measurement campaigns and proofreading the manuscript, and A. Schnegg and D. Ponwitz for support at the THz-beamline at Bessy II.

\end{acknowledgments}




\begin{thebibliography}{9}\label{sec:literature}%

\bibitem{Glatter1995} O. Glatter and O. Kratky, \emph{Small Angle X-ray Scattering} (Academic Press, Boston, 1982).

\bibitem{Xu2002} R. Xu, \emph{Particle Characterization: Light Scattering Methods} (Kluwer Springer Netherlands, Heidelberg, 2002).

\bibitem{Brown1996} W. Brown, \emph{Light Scattering: Principles and development} (Clarendon Press, Oxford, 1996).

\bibitem{Hulst1981} H. C. van de Hulst, \emph{Light Scattering by Small Particles} (Dover Publications, New York, 1981).

\bibitem{Bohren1983} C. F. Bohren and D. R. Huffman, \emph{Absorption and Scattering of Light by Small Particles} (John Wiley \& Sons, Weinheim, 1983).

\bibitem{Born1999} M. Born and E. Wolf, \emph{Principles of optics, seventh (expanded) edition} (Cambrige University Press, Cambridge, 1999).

\bibitem{Ishimaru1999} A. Ishimaru, \emph{Wave Propagation and Scattering in Random Media} (John Wiley \& Sons, Weinheim, 1999).

\bibitem{Mishchenko2014} M. I. Mishchenko, \emph{Electromagnetic Scattering by Particles and Particle Groups} (Cambrige University Press, Cambridge, 2014).

\bibitem{Mishchenko2000} M. I. Mishchenko,  J. W. Hovenier and L. D. Travis \emph{Light Scattering by Nonspherical Particles} (Academic Press, San Diego, 2000).

\bibitem{Born2015} P. Born, K. Holldack, and M. Sperl, \emph{Granul. Matter} \textbf{17}, 531 (2015).

\bibitem{Waseda1980} Y. Waseda, \emph{The structure of non-crystalline materials: Liquids and amorphous solids} (McGraw-Hill International Book Co, New York , 1980).

\bibitem{Hansen2005} J.-P. Hansen and I. R. McDonald, \emph{Theory of simple liquids} (Academic Press, London, 2005).

\bibitem{Fraden1990} S. Fraden and G. Maret, \emph{Phys. Rev. Lett.} \textbf{65} 512 (1990).

\bibitem{Auger2011} J.-C. Auger and B. Stout, \emph{J. Coatings Technol. Res.} \textbf{9}, 287 (2011).

\bibitem{Born2014} P. Born, N. Rothbart, M. Sperl, and H.-W. H\"{u}bers, \emph{Europhysics Lett.} \textbf{106}, 48006 (2014).

\bibitem{Medebach2007} M. Medebach, C. Moitzi, N. Freiberger, and O. Glatter, \emph{J. Colloid Interface Sci.} \textbf{305}, 88 (2007).

\bibitem{Kristensson2015} G. Kristensson, \emph{J. Quant. Spectrosc. Radiat. Transf.} \textbf{164}, 97 (2015).

\bibitem{Weitz1993} D. A. Weitz and D. J. Pine, \emph{Diffusing wave spectroscopy}, in: \emph{Dynamic Light Scattering: The Method and Some Applications}, W. Brown, Ed. (Oxford University Press, Oxford, 1993).

\bibitem{Rojas2004} L. Rojas-Ochoa, J. Mendez-Alcaraz, J. S\'{a}enz, P. Schurtenberger, and F. Scheffold, \emph{Phys. Rev. Lett.} \textbf{93}, 73903 (2004).

\bibitem{Kaplan1994} P. D. Kaplan, A. D. Dinsmore, A. G. Yodh, and D. J. Pine, \emph{Phys. Rev. E} \textbf{50}, 4827 (1994).

\bibitem{Nisbet2015} A. G. A. Nisbet, G. Beutier, F. Fabrizi, B. Moser and S. P. Collins, \emph{Acta Cryst.} \textbf{A71}, 20 (2015).

\bibitem{Skipetrov2016} S. E. Skipetrov and J. H. Page, \emph{New J. Phys.} \textbf{18}, 21001 (2016).

\bibitem{Sperling2016} T. Sperling, L. Schertel, M. Ackermann, G. J. Aubry, C. M. Aegerter and G. Maret, \emph{New J. Phys.} \textbf{18}, 013039 (2016).

\bibitem{Petrova2009} E. V. Petrova, V. P. Tishkovets, and K. Jockers, \emph{Sol. Syst. Res.} \textbf{43}, 100 (2009).

\bibitem{Schaefer2012} J. Sch\"afer, S. Lee, and A. Kienle, \emph{J. Quant. Spectrosc. Radiat. Transf.} \textbf{113}, 2113 (2012).

\bibitem{Rezvani2015} R. Rezvani Naraghi, S. Sukhov, J. J. S\'{a}enz, and A. Dogariu, \emph{Phys. Rev. Lett.} \emph{115}, 203903 (2015).

\bibitem{Froufe2016} L. S. Froufe-P\'{e}rez, M. Engel, P. F. Damasceno, N. Muller, J. Haberko, S. C. Glotzer, and F. Scheffold, \emph{Phys. Rev. Lett.} \textbf{117}, 53902 2016.

\bibitem{Hapke1993} B. Hapke, \emph{Theory of Reflectance and Emittance Spectroscopy} (Cambrige University Press, Cambridge, 1993).

\bibitem{Kolokova2001} L. Kolokolova and B. A. S. Gustafson, \emph{J. Quant. Spectrosc. Radiat. Transf.} \textbf{70}, 611 (2001).

\bibitem{Tsang2000} L. Tsang, C.-T Chen, A. T. C. Chang, J. Guo, and K.-H. Ding, \emph{Radio Sci.} \textbf{35}, 731 (2000).

\bibitem{Feigin1987} L. A. Feigin and D. I. Svergun, \emph{Structure Analysis by Small-Angle X-ray and Neutron Scattering} (Plenum Press, New York, 1987)

\bibitem{Taday2004} P. F. Taday, \emph{Philos. Trans. A. Math. Phys. Eng. Sci.} \textbf{362}, 351 (2004).

\bibitem{Bandyopadhyay2007} A. Bandyopadhyay, A. Sengupta, R. B. Barat, D. E. Gary, J. F. Federici, M. Chen, and D. B. Tanner, \emph{Int. J. Infrared Millimeter Waves} \textbf{28}, 969 (2007).

\bibitem{Zurk2008} L. M. Zurk, G. Sundberg, S. Schecklman, Z. Zhou, A. Chen, and E. I. Thorsos, \emph{Terahertz for Military and Security Applications VI,} \textbf{6949}, 694907 (2008).

\bibitem{Shen2008} Y. C. Shen, P. F. Taday, and M. Pepper, \emph{Appl. Phys. Lett.} \textbf{92}, 51103 (2008).

\bibitem{Kaushik2012} M. Kaushik, B. W.-H. Ng, B. M. Fischer, and D. Abbott, \emph{IEEE Photonics Technol. Lett.} \textbf{24}, 155 (2012).

\bibitem{May2009} R. K. May, M. Evans, S. Zhong, R. Clarkson, Y. Shen, L. F. Gladden, and J. A. Zeitler, \emph{Real-time in situ measurement of particle size in flowing powders by terahertz time-domain spectroscopy} in \emph{2009 34th Int. Conf. Infr., Millim., THz Waves}, 1 (2009).

\bibitem{Pearce2004} J. Pearce, Z. Jian, and D. M. Mittleman, \emph{Philos. Trans. A. Math. Phys. Eng. Sci.} \textbf{362}, 301 (2004).

\bibitem{Pearce2001} J. Pearce and D. M. Mittleman, \emph{Opt. Lett.} \textbf{26}, 2002 (2001).

\bibitem{Gerthsen2010} D. Meschede, \emph{Gerthsen Physik} (Springer, Heidelberg, 2010).

\bibitem{Takagi2004} K. Takagi, K. Seno, and A. Kawasaki, \emph{Appl. Phys. Lett.} \textbf{85}, 3681 (2004).

\bibitem{Takagi2010} K. Takagi, M. Omote, and A. Kawasaki, \emph{J. Micromechanics Microengineering} \textbf{20}, 35032 (2010).

\bibitem{Brundermann2011} E. Br\"{u}ndermann, H.-W- H\"{u}bers and M. Kimmitt, \emph{Terhertz Techniques}(Springer, Heidelberg, 2011).

\bibitem{Peiponen2013} K.-E. Peiponen, J. A. Zeitler and M. Kuwata-Gonokami, \emph{Terahertz Spectroscopy and Imaging} (Springer, Berlin, 2013).

\bibitem{Naftaly2015} N. Naftaly, \emph{Terhertz metrology}(Artech House, Boston, 2015).

\bibitem{Piesiewicz2007} R. Piesiewicz, C. Jansen, S. Wietzke, D. Mittleman, M. Koch, and T. Kürner, \emph{Int. J. Infrared Millimeter Waves} \textbf{28}, 363 (2007).

\bibitem{Folks2007}W. R. Folks, S. K. Pandey, and G. Boreman, \emph{Opt. Terahertz Sci. Tech.} \textbf{2}, 10 (2007).

\bibitem{Cunningham2011} P. D. Cunningham, N. N. Valdes, F. A. Vallejo, L. M. Hayden, B. Polishak, X.-H. Zhou, J. Luo, A. K.-Y. Jen, J. C. Williams, and R. J. Twieg, \emph{J. Appl. Phys.} \textbf{109}, 43505 (2011).

\bibitem{Richter2010} H. Richter, M. Greiner-B\"{a}r, S. Pavlov, A. D. Semenov, M. Wienold, L. Schrottke, M. Giehler, R. Hey, H. T. Grahn, and H. W. H\"{u}bers, \emph{Opt. Express} \textbf{18}, 5890 (2010).

\bibitem{Holldack2016} K. Holldack and A. Schnegg, \emph{Journal of large-scale research facilities} \textbf{2}, A51 (2016). http://dx.doi.org/10.17815/jlsrf-2-74

\bibitem{Leutz1996} W. Leutz and J. Ricka, \emph{Opt. Commun.} \textbf{126}, 260 (1996).

\bibitem{Abo2003} M. Abo-Bakr, J. Feikes, K. Holldack, P. Kuske, W. Peatman, U. Schade, G. W\"{u}stefeld, and H.-W. H\"{u}bers, \emph{Phys. Rev. Lett.} \textbf{90}, 94801 (2003).

\bibitem{Withaya2006} W. Withayachumnankul, B. Fergusson, T. Rainsford, S. Mickan, and D. Abbot, \emph{Fluct. Noise Lett.} \textbf{6}, L227 (2006).

\bibitem{Holldack2007} K. Holldack and D. Ponwitz, \emph{AIP Conference Proceedings} \textbf{879}, 599 (2007).
\end{thebibliography}
\end{document}